\documentstyle[11pt,amssymb,epsf]{article}

\def\ben{\begin{equation}}
\def\een{\end{equation}}

  \let\n=\nu

\let\C=\Chi

\def\nn{\nonumber} \def\bd{\begin{document}} \def\ed{\end{document}}
\def\ds{\documentstyle} \let\fr=\frac \let\bl=\bigl \let\br=\bigr
\let\Br=\Bigr \let\Bl=\Bigl
\let\bm=\bibitem
\let\na=\nabla
\let\pa=\partial \let\ov=\overline
\newcommand{\be}{\begin{equation}}
\newcommand{\ee}{\end{equation}}
\def\ba{\begin{array}}
\def\ea{\end{array}}
\def\ft#1#2{{\textstyle{{\scriptstyle #1}\over {\scriptstyle #2}}}}
\def\fft#1#2{{#1 \over #2}}
\def\del{\partial}
\def\vp{\varphi}
\def\sst#1{{\scriptscriptstyle #1}}
\def\oneone{\rlap 1\mkern4mu{\rm l}}
\def\td{\tilde}
\def\wtd{\widetilde}
\def\ie{\rm i.e.\ }
\def\dalemb#1#2{{\vbox{\hrule height .#2pt
        \hbox{\vrule width.#2pt height#1pt \kern#1pt
                \vrule width.#2pt}
        \hrule height.#2pt}}}
\def\square{\mathord{\dalemb{6.8}{7}\hbox{\hskip1pt}}}
\newcommand{\ho}[1]{$\, ^{#1}$}
\newcommand{\hoch}[1]{$\, ^{#1}$}
\newcommand{\bea}{\begin{eqnarray}}
\newcommand{\eea}{\end{eqnarray}}
\newcommand{\ra}{\rightarrow}
\newcommand{\lra}{\longrightarrow}
\newcommand{\Lra}{\Leftrightarrow}
\newcommand{\ap}{\alpha^\prime}
\newcommand{\bp}{\tilde \beta^\prime}
\newcommand{\tr}{{\rm tr} }
\newcommand{\Tr}{{\rm Tr} }
\def\0{{\sst{(0)}}}
\def\1{{\sst{(1)}}}
\def\2{{\sst{(2)}}}
\def\3{{\sst{(3)}}}
\def\4{{\sst{(4)}}}
\def\5{{\sst{(5)}}}
\def\6{{\sst{(6)}}}
\def\7{{\sst{(7)}}}
\def\8{{\sst{(8)}}}
\def\n{{\sst{(n)}}}
\def\cA{{{\cal A}}}
\def\cF{{{\cal F}}}
\def\tV{\widetilde V}
\def\tW{\widetilde W}
\def\tH{\widetilde H}
\def\tE{\widetilde E}
\def\tF{\widetilde F}
\def\tA{\widetilde A}
\def\im{{{\rm i}}}
\def\tY{{{\wtd Y}}}
\def\ep{{\epsilon}}
\def\vep{{\varepsilon}}
\def\R{\rlap{\rm I}\mkern3mu{\rm R}}
\def\bD{{{\bar D}}}

\def\R{\rlap{\rm I}\mkern3mu{\rm R}}
\def\bD{{{\bar D}}}
\def\R{{{\Bbb R}}}
\def\C{{{\Bbb C}}}
\def\H{{{\Bbb H}}}
\def\CP{{{\Bbb C}{\Bbb P}}}
\def\RP{{{\Bbb R}{\Bbb P}}}
\def\Z{{{\Bbb Z}}}
\def\bA{{{\Bbb A}}}
\def\bB{{{\Bbb B}}}
\def\bC{{{\Bbb C}}}
\def\bD{{{\Bbb D}}}
\def\bZ{{{\Bbb Z}}}
\def\Re{{{\frak{Re}}}}
\def\cosec{{\,\hbox{cosec}\,}}

\newcommand{\tamphys}{\it Center for Theoretical Physics,
Texas A\&M University, College Station, TX 77843, USA}
\newcommand{\umich}{\it Michigan Center for Theoretical Physics,
University of Michigan\\ Ann Arbor, MI 48109, USA}
\newcommand{\upenn}{\it Department of Physics and Astronomy,
University of Pennsylvania\\ Philadelphia,  PA 19104, USA}
\newcommand{\SISSA}{\it  SISSA-ISAS and INFN, Sezione di Trieste\\
Via Beirut 2-4, I-34013, Trieste, Italy}

\newcommand{\ihp}{\it Institut Henri Poincar\'e\\
  11 rue Pierre et Marie Curie, F 75231 Paris Cedex 05}

\newcommand{\damtp}{\it DAMTP, Centre for Mathematical Sciences,
 Cambridge University\\ Wilberforce Road, Cambridge CB3 OWA, UK}
\newcommand{\itp}{\it Institute for Theoretical Physics, University of
California\\ Santa Barbara, CA 93106, USA}

\newcommand{\auth}{M. Cveti\v{c}\hoch{\dagger}, G.W. Gibbons\hoch{\sharp},
H. L\"u\hoch{\star} and C.N. Pope\hoch{\ddagger}}

\begin{document}
\begin{flushright}
\hfill{DAMTP-2001-107}\ \ \ {CTP TAMU-28/33}\ \ \ {UPR-972-T}\ \ \
{MCTP-01-61}\\
{December 2001}\\
{hep-th/0112098}
\end{flushright}


\begin{center}
{ \large {\bf M-theory Conifolds}}

\vspace{5pt}
\auth

\vspace{3pt}
{\hoch{\dagger}\upenn}

\vspace{3pt}


\vspace{3pt}
{\hoch{\sharp}\damtp}

\vspace{3pt}
{\hoch{\star}\umich}

\vspace{3pt}
{\hoch{\ddagger}\tamphys}

\vspace{3pt}

\underline{ABSTRACT}
\end{center}

  Seven-manifolds of $G_2$ holonomy provide a bridge between M-theory
and string theory, via Kaluza-Klein reduction to Calabi-Yau
six-manifolds.  We find first-order equations for a new family of
$G_2$ metrics $\bD_7$, with $S^3\times S^3$ principal orbits.  These
are related at weak string coupling to the resolved conifold,
paralleling earlier examples $\bB_7$ that are related to the deformed
conifold, allowing a deeper study of topology change and mirror
symmetry in M-theory.  The $\bD_7$ metrics' non-trivial parameter
characterises the squashing of an $S^3$ bolt, which limits to $S^2$ at
weak coupling.  In general the $\bD_7$ metrics are asymptotically
locally conical, with a nowhere-singular circle action.

{\vfill\leftline{}\vfill
\vskip 5pt
\footnoterule
{\footnotesize \hoch{\dagger} Research supported in part by DOE grant
DE-FG02-95ER40893 and NATO grant 976951. \vskip -12pt} \vskip 14pt
{\footnotesize \hoch{\star} Research supported in full by DOE grant
DE-FG02-95ER40899 \vskip -12pt} \vskip 14pt
{\footnotesize  \hoch{\ddagger} Research supported in part by DOE
grant DE-FG03-95ER40917.\vskip  -12pt}}

\pagebreak
\setcounter{page}{1}

\vfill\eject

    Calabi-Yau manifolds, both compact and non-compact, singular and
non-singular, have long been studied because of their significance for
string theory, since they provide a way of obtaining ${\cal N}=1$
supersymmetry in four dimensions.  The principal non-compact example
is the singular conifold, and its smoothed-out versions, namely the
resolved conifold and the deformed conifold \cite{candel}.  The
singular apex of the cone over $T^{1,1}= (S^3\times S^3)/S^1$ is blown
up to a smooth 2-sphere in the former, and to a smooth 3-sphere in the
latter.  These minimal (calibrated) surfaces are supersymmetric cycles
over which D-branes may be wrapped.  If one considers a sequence of
smooth models in which the cycles shrink to zero, one obtains enhanced
gauge symmetry at the conifold point, the resolved and deformed
conifolds being related by mirror symmetry \cite{smokeand}.  Studying
this process has led to an understanding of topology change in quantum
gravity \cite{strom}.  

   With the advent of M-theory, it has become important to consider
the lifts of these 6-manifolds with holonomy $SU(3)$ to seven
dimensions and holonomy $G_2$ \cite{achvaf,atmava,atiwit}, in order to
set the four-dimensional ${\cal N}=1$ theories in an M-theory context.
The seven-dimensional and six-dimensional manifolds are related by
Kaluza-Klein reduction on a circle, whose variable length $R$ is
related to the coupling constant $g$ of type IIA string theory by $R
\propto g^{2/3}$.  Thus we seek asymptotically locally conical (ALC)
$G_2$ manifolds, for which the size $R$ of the circle tends to a
constant at infinity.  In the case that $R$ is everywhere constant,
and the associated Kaluza-Klein vector field vanishes, the
six-dimensional manifold is an exact Calabi-Yau space.  If $R$ varies,
it will be an approximate Calabi-Yau space.  This approximation will
be good everywhere if the coupling constant $g$, or, equivalently, the
radius $R$, never vanishes and is slowly varying.  One may show on
general grounds that it is never larger than its value at infinity.

   Since the principal orbits of the the smoothed-out conifold are
$T^{1,1}$, it follows that the principal orbits of the associated
seven-dimensional $G_2$ metrics will be a $U(1)$ bundle over
$T^{1,1}$, which is in fact $S^3\times S^3$.  Very few examples of
cohomogeneity one $G_2$ metrics can arise \cite{cleyswan}, and in
fact the only explicitly-known examples have principal orbits that are
$\CP^3$, the flag manifold $SU(3)/(U(1)\times U(1))$, and $S^3\times
S^3$.  Asymptotically conical (AC) metrics are known for all three
cases \cite{brysal,gibpagpop}, but only the $S^3\times S^3$ case has
enough freedom to permit ALC metrics of cohomogeneity one to arise.

    In previous work \cite{slumpy}, we presented complete non-singular
$G_2$ metrics, which we denoted by $\bC_7$, for which the coupling
constant varied in such a finite positive interval.  The associated
Calabi-Yau space is the Ricci-flat K\"ahler metric on a complex line
bundle over $S^2\times S^2$ \cite{berber,pagpop}.  Other work has
provided $G_2$ metrics $\bB_7$ associated with the deformed conifold
Calabi-Yau space \cite{bragogugu,cglp2}.  However, in this case the
radius $R$ vanishes on an $S^3$ supersymmetric (calibrated) cycle in
the interior.  The purpose of this present letter is to extend the
picture by providing a new class of complete non-singular $G_2$
metrics, which we denote by $\bD_7$, whose associated Calabi-Yau
manifold is the resolved conifold.  In this case, as in the $\bC_7$ metrics,
the coupling constant never vanishes.  The new metrics provide a
unifying link between the deformed and resolved conifolds, via strong
coupling and M-theory.

     The metrics are invariant under the action of $SU(2)\times SU(2)$,
with left-invariant 1-forms $\sigma_i$ and $\Sigma_i$.  The metric ansatz
is
\be
ds_7^2 = dt^2 + a^2\, ((\Sigma_1+ g\, \sigma_1)^2 +
(\Sigma_2+ g\,\sigma_2)^2) + b^2\, (\sigma_1^2 + \sigma_2^2) +
c^2 (\Sigma_3 +g_3\, \sigma_3)^2 + f^2 \sigma_3^2\,,\label{d7ans}
\ee
where $a$, $b$, $c$, $f$, $g$ and $g_3$ are functions only of the
radial variable $t$.  If we write $\sigma_i$ in terms of Euler angles,
with $\sigma_1+\im\, \sigma_2= e^{-\im\, \psi}\, (d\theta + \im\,
\sin\theta\, d\phi)$, $\sigma_3=d\psi + \cos\theta\, d\phi$, and
similar expressions using tilded Euler angles for $\Sigma_i$, then the
M-theory circle is generated by $\psi\longrightarrow \psi + k$,
$\wtd\psi\longrightarrow \wtd\psi + k$, where $k$ is a constant.  This
$U(1)$ diagonal subgroup of the right translations\footnote{The metric
ansatz (\ref{d7ans}) is a specialisation of a nine-function ansatz
introduced in \cite{cglp2}, in which the metric functions for the
$i=1$ and $i=2$ directions in the two $SU(2)$ groups are set equal.
The diagonal $U(1)$ subgroup of the $SU(2)$ right-translations becomes
an isometry, as is needed for Kaluza-Klein reduction, under this
specialisation.}  is generated by the Killing vector $K=\del/\del\psi
+ \del/\del\wtd\psi$.  The orbits of $SU(2)\times SU(2)$ are
generically six-dimensional.  In our solutions, the orbits collapse in
the interior to a 3-sphere, which in general has a squashed rather
than round $SU(2)$-invariant metric.  The degenerate orbit is known as
a bolt; it is a minimal surface and a supersymmetric (associative)
3-cycle.

  The metric will have $G_2$ holonomy, and thus will also be Ricci flat,
if it admits a closed and co-closed associative 3-form
\be
\Phi_\3 =e^0\, e^3\, e^6 +e^1\, e^2\, e^6 - e^4\, e^5\, e^6
+ e^0\, e^1\, e^4 + e^0\, e^2\, e^5 - e^1\,
e^3\, e^5 + e^2\, e^3\, e^4\,,\label{assoc}
\ee
where the vielbein is given by $e^0=dt$, $e^1=a\, (\Sigma_1 +g\,
\sigma_1)$, $e^2=a\, (\Sigma_2 +g\, \sigma_2)$, $e^3=c\, (\Sigma_2
+g_3\, \sigma_2)$, $e^4=b\, \sigma_1$, $e^5=b\, \sigma_2$ and $e^6=f\,
\sigma_3$. The closure and co-closure implies the algebraic
constraints
\be
g=-\fft{a\, f}{2b\, c}\,,\qquad
g_3=-1 + 2g^2\,,\label{alg}
\ee
together with the first-order equations
\bea
\dot a &=& -\fft{c}{2a} + \fft{a^5\, f^2}{8b^4\, c^3}\,,\qquad
\qquad\qquad\quad\
\dot b = -\fft{c}{2b} - \fft{a^2\, (a^2-3c^2)\, f^2}{8b^3\, c^3}\,,\nn\\
\dot c &=& -1 +\fft{c^2}{2a^2} + \fft{c^2}{2b^2}
-\fft{3 a^2\, f^2}{8b^4}\,,\qquad
\dot f = -\fft{a^4\, f^3}{4b^4\,  c^3}\,.\label{newfo}
\eea

    Using the closure and co-closure conditions has reduced the
Einstein equations, which are of second order and extremely
complicated, to a manageable first-order set involving just the four
functions $a$, $b$, $c$ and $f$.  One can check that the equations are
a consistent truncation of the second-order Einstein equations for the
more general nine-function ansatz that was given in \cite{cglp2}.  It
should be emphasised that although the equations here have reduced to
a four-function first-order system, the ansatz is inequivalent to the
four-function ansatz introduced in \cite{bragogugu}.  In particular
the metric ansatz in \cite{bragogugu} admits a $Z_2$ symmetry under
which the $\sigma_i$ and $\Sigma_i$ are interchanged and the
associative 3-form changes sign, whilst our metric ansatz
(\ref{d7ans}) does not have this symmetry.\footnote{We understand that
S. Gukov, K. Saraikin and N. Volovitch are also considering ans\"atze
that break the $Z_2$ symmetry \cite{gukov}.}

   We can find a regular series expansion for the situation where both
$a$ and $c$ go to zero at short distance.  Substituting the Taylor
expansions for the four functions $a$, $b$, $c$ and $f$ into
(\ref{newfo}), we find
\bea
a &=& \fft{t}{2} -\fft{(q^2+2)\, t^3}{288} -
   \fft{(31q^4 -29 q^2 -74)\, t^5}{69120} + \cdots\,,\nn\\
b&=& 1 - \fft{(q^2-2)\, t^2}{16} - \fft{(11q^4-21q^2+13)\, t^4}{1152} +\cdots
\,,\nn\\
c&=& -\fft{t}{2} - \fft{(5q^2-8)\, t^3}{288} -
       \fft{(157 q^4 -353 q^2 +232)\, t^5}{34560} +\cdots\,,\nn\\
f&=& q + \fft{q^3\, t^2}{16} + \fft{q^3\, (11q^2-14)\, t^4}{1152}
  +\cdots\,,
\eea
where, without loss of generality, we have set the scale size so that
$b=1$ on the $S^3$ bolt at $t=0$.  The parameter $q$ is free, and
characterises the squashing of the $S^3$ bolt along its $U(1)$ fibres
over the unit $S^2$.  By studying the equations numerically, using the
short-distance Taylor expansion to set initial data just outside the
bolt, we find that there is a regular asymptotically conical (AC)
solution when $q=1$, and that there are regular ALC solutions for any
$q$ in the interval $0<q<1$.  In fact the AC solution at $q=1$ is the
well-known $G_2$ metric on the spin bundle of $S^3$, found in
\cite{brysal,gibpagpop}.  (One can easily derive this analytically
from (\ref{newfo}), by noting that it corresponds to the consistent
truncation $c=-a$, $f=b$.)  The ALC solutions with the non-trivial
parameter $0<q<1$ are new, and we shall denote them by $\bD_7$.  They
exhibit the unusual phenomenon of admitting a supersymmetric
Lagrangian 3-manifold (the bolt) that is not Einstein.  The metric function
$f$ tends to a constant at infinity, while the remaining functions $a$, $b$
and $c$ grow linearly with $t$; in fact $a$, $b$ and $c$ satisfy the
first-order equations governing the Ricci-flat K\"ahler resolved conifold
asymptotically at large distance.  One can see from (\ref{d7ans}) that
the $U(1)$ Killing vector $K=\del/\del\psi+\del/\del\wtd\psi$ has length
given by $|K|^2= f^2 + c^2\, (1+g_3)^2$, and so it follows that its
length is nowhere infinite or zero.  It ranges from a minimum value $|K|=q$
at short distance to the asymptotic value $|K|=f_\infty$ at infinity.

     It may well be that the system (\ref{newfo}) is completely
integrable,\footnote{By contrast, it is expected that the
second-order Einstein equations for Ricci flatness are of the type
that would give rise to chaotic behaviour \cite{damhen}.} although we
have not yet succeeded in finding the general solution to the
first-order equations.  In a somewhat analogous situation in eight
dimensions, we did find the general solution to the first-order
equations for an ansatz for ALC metrics of Spin(7) holonomy
\cite{cglp3}.  In that case, the first-order equations could be
reduced to an autonomous third-order equation, whose general solution
could be given in term of hypergeometric functions.  In the present
case, we can again reduce the first-order equations to an autonomous
third-order equation for $G\equiv g^2$:
\bea
&&[( - 6\,G^{2} + 2 G)\, G'^{2} - 4( 7G^{3} -
 2G^{2})\,  G' + 8 G^{3} - 32 G^{4}] \,G'''\nn\\
&&
+[(3\,G + 1)\,(\,
G')^{3} + 6 (14 G^{2} - 3G)
\, G'^{2} - 4(9 G^{2} - 31 G^{3})\,G' + 8 G^{3} - 32 G^{4}] \,G''
\nn\\
&&+ [6(3 G^{2} - G)\,G' - 12G^{2}
 + 32G^{3}]\, {G''}^{2}
+ 2(3G + 1)\,G'^{4} - 20(G - 6G^{2})\,G'^{3} \nonumber\\
&& -8(7G^{2} - 29 G^{3})\,G'^{2} -16 (G^{3} + 4G^{4})\,G' =0\,,
\eea
with $A\equiv c^2/a^2=1+ G'/(2G)$, $c^2/b^2=(A'+2A^2 -2A)/(G+3G\,
A-A)$, and $a\, b\, c=e^{\rho}$.  The primes denote derivatives with
respect to the new radial variable $\rho$, defined by $dt=-c\,
d\rho$. We have found the following new explicit solution,
\bea
 ds^2&=& h^{-1/3}\, dr^2+ \ft16 r^2\, h^{-1/3}\, 
[(\Sigma_1+{k\over r}\sigma_1)^2+(\Sigma_2+{k\over
r}\sigma_2)^2]\nn\\
 &&+\ft19 r^2\, h^{-1/3}\, [\Sigma_3+(-1+\fft{2k^2}{r^2})\,\sigma_3]^2+
\ft16r^2\, h^{2/3}\, (\sigma_1^2+\sigma_2^2)+\ft49 k^2\, 
h^{2/3}\, \sigma_3^2\,,\label{singmet}  
\eea
where $h\equiv 1-9k^2/(2r^2)$.  Unlike the smooth $\bD_7$ metrics that we
have found numerically, (\ref{singmet}) has a curvature singularity at
$r^2=9k^2/2$.  

   It is useful to summarise some known results for $G_2$ metrics with
$S^3\times S^3$ principal orbits in the form of a table.

\bigskip\bigskip

\centerline{
\begin{tabular}{|c|c|c|c|c|c|}\hline
$G_2$ Metric & Calabi-Yau  & Bolt & AC limit & Susy cycle? & $\Z_2$ sym?
\\ \hline\hline
$\bB_7$ & Deformed conifold & $S^3_1$
& $\R^4\times S^3$ & Yes & Yes\\ \hline
$\bC_7$  & $\C\ltimes (S^2\times S^2)$
& $T^{1,1}_q$ & $\sim\R^4\times S^3$ & No & Yes\\ \hline
$\bD_7$ & Resolved conifold & $S^3_q$
& $\R^4\times S^3$ & Yes & No \\ \hline
\end{tabular}}
\bigskip

\centerline{Table 1: The three families of $G_2$ solutions}
\bigskip\bigskip

   We are including three families of complete non-singular solutions
here, each of which has a non-trivial parameter.  At one end of the
parameter range the metric is asymptotically conical.  For the $\bB_7$
and $\bD_7$ cases, this AC metric is precisely the one found in
\cite{brysal,gibpagpop}, on the spin bundle of $S^3$.  Since the
bundle is trivial, we are denoting this AC metric simply by
$\R^4\times S^3$.  In the case of the $\bC_7$ metrics \cite{slumpy},
the limiting AC member of the family approaches the form of the AC
metric of \cite{brysal,gibpagpop} at large distance, but is quite
different at short distance, since it instead has the topology of an
$\R^2$ bundle over $T^{1,1}$.  For the $\bC_7$ metrics, and our new
$\bD_7$ metrics, the non-trivial parameter in the metrics
characterises the degree of squashing of the $T^{1,1}$ or $S^3$ bolt
respectively, as denoted by the subscripts $q$ on $T^{1,1}_q$ and
$S^3_q$ in the Table.  By contrast, for the $\bB_7$ metrics the $S^3$
bolt is always round (denoted by the subscript ``1'' on $S^3_1$), and
the non-trivial parameter instead characterises ``velocities'' of the
metric functions as one moves outwards from the bolt \cite{cglp2}.
(An explicit solution for one specific value of the non-trivial
parameter was obtained in \cite{bragogugu}.)

   As the non-trivial parameter in the ALC metric is reduced from its
AC limiting value, a circle ``splits off'' and stabilises its length
when one moves out sufficiently far.  The geometry is that of a twisted
circle bundle over a six-dimensional AC metric.  At the lower limit of
the parameter range the radius of the circle at infinity becomes
vanishingly small.  If one performs an appropriate counterbalancing
rescaling of the circle coordinate, the Kaluza-Klein vector describing
the twist vanishes in the limit and one obtains the Gromov-Hausdorff
limit which is just the direct product of $S^1$ times a Ricci-flat
Calabi-Yau six-metric.  Thus the Gromov-Hausdorff limit may be
identified with the weak coupling limit in this case.
These metrics are listed in the second column of the Table.
The metric $\C\ltimes (S^2\times S^2)$ denotes the Ricci-flat K\"ahler metric
on the complex line bundle over $S^2\times S^2$ that was constructed in
\cite{berber,pagpop}.

   The $\bB_7$ and $\bD_7$ metrics provide a seven-dimensional link
between the six-dimensional deformed and resolved conifolds.  This can
be seen from the fact that both the $\bB_7$ and $\bD_7$ families of
metrics are encompassed by the ansatz (\ref{d7ans}).  They satisfy two
different systems of first-order equations that are each consistent
truncations of the same system of six second-order Ricci-flat
equations.  Each of the $\bB_7$ and $\bD_7$ families has a continuous
non-trivial modulus parameter, with each family having the {\it same} 
AC metric at one end of the parameter range, whilst at the other end
of the range the $\bB_7$ and $\bD_7$ metrics approach $S^1$ times the
deformed conifold and the resolved conifold respectively.  This
implies that the two weakly coupled IIA string theory backgrounds
using the deformed and the resolved conifolds are related via strong
coupling and eleven dimensions.

    An important issue for future work is the phenomenologically
central question of chiral fermions localised at isolated
singularities \cite{cveshiura,witten,achwit}. Physically, these can
arise in M-theory from massless states associated to membranes wrapped
around vanishing cycles. Mathematically, they correspond to solutions
of the massless Dirac equation in the M-theory background.  The
process of localisation is as yet imperfectly understood.  What is
needed is explicit metrics permitting explicit calculations.  Our
metrics are certainly sufficiently simple for this purpose.  What
requires further investigation is whether one can model the
appropriate co-dimension seven singularities using them.

\section*{Acknowledgements}

   We are grateful to Gary Shiu for helpful discussions.  G.W.G.,
H.L. and C.N.P. are grateful to UPenn for support and hospitality
during the course of this work.

\end{document}